\documentclass[aps,pre,twocolumn,showpacs,10pt]{revtex4-1}

\usepackage{graphicx} 
\usepackage{color}
\definecolor{Green}{rgb}{0,0.5,0}

\usepackage[english]{babel}
\usepackage[utf8]{inputenc}

\usepackage{amssymb}

\usepackage{textcomp}

\begin{document}

\title[Characterizing wave-functions in graphene nanodevices]{Characterizing wave functions in graphene nanodevices: electronic transport through ultrashort graphene constrictions on a boron nitride substrate}
\author{D. Bischoff$^1$}
\email{dominikb@phys.ethz.ch}
\author{F. Libisch$^2$}
\author{J. Burgdörfer$^2$}
\author{T. Ihn$^1$}
\author{K. Ensslin$^1$}
\affiliation{$^1$ Solid State Physics Laboratory, ETH Zurich, 8093 Zurich, Switzerland\\ $^2$ Institute for Theoretical Physics, Vienna University of Technology, A-1040 Vienna, Austria}
\date{\today}



\begin{abstract}
We present electronic transport measurements through short and narrow ($30\times30\,\mathrm{nm}$) single layer graphene constrictions on a hexagonal boron nitride substrate. While the general observation of Coulomb-blockade is compatible with earlier work, the details are not: we show that the area on which charge is localized can be significantly larger than the area of the constriction, suggesting that the localized states responsible for Coulomb-blockade leak out into the graphene bulk. The high bulk mobility of graphene on hexagonal boron nitride, however, seems not consistent with the short bulk localization length required to see Coulomb-blockade. To explain these findings, charge must instead be primarily localized along the imperfect edges of the devices and extend along the edge outside of the constriction. In order to better understand the mechanisms, we compare the experimental findings with tight-binding simulations of such constrictions with disordered edges. Finally we discuss previous experiments in the light of these new findings.
\end{abstract}



\maketitle


\section*{Introduction}
Right after graphene became available for experiments~\cite{Novoselov2004}, narrow graphene stripes called either nanoribbons or (nano)constrictions were investigated in great detail. As graphene nanoconstrictions are the simplest and most basic building blocks for other graphene nano\-devices, it is crucial to understand their properties in detail before effects in more elaborate devices can be investigated and understood successfully. Furthermore, early theoretical work predicted that a bandgap dependent on the ribbon width~\cite{Nakada1996,Ezawa2006} opens, which would allow to tune the material parameters by simply changing the geometry. Graphene nanoribbons were fabricated with a large variety of different methods, where etching by plasma (e.g. Refs.~\cite{Han2007,Chen2007,Liu2009,Molitor2009,Todd2009,Bai2010,Oostinga2010,Candini2011,Ribeiro2011,Terres2011,Behnam2012,Duerr2012,Hwang2012,Minke2012}) and unzipping of carbon nanotubes (e.g. Refs.~\cite{Li2008,Kosynkin2009}) are the two most widely used techniques. A comprehensive list of fabrication methods, different findings and measurement techniques can be found in the  supplementary materials~\cite{Supplementary}. Devices fabricated by these different methods will likely differ on a microscopic level in terms of edge structure, contaminants and coupling to the substrate. 

There is a number of interpretations of the data from different experiments. For the experiment presented in this paper, it is unlikely that the bandgap expected from band-structure calculations of perfect graphene nanoribbons~\cite{Nakada1996,Ezawa2006} plays a significant role in explaining the data because the ribbons are too wide and the edges are too disordered. In addition, the crystallographic orientation is generally unknown. In various experiments clear signatures of Coulomb-blockade (see e.g. Refs.~\cite{Liu2009,Molitor2009,Todd2009,Bai2010,Han2010,Oostinga2010,Shimizu2011,Candini2011,Terres2011,Wang2011,Behnam2012}) together with a temperature dependence compatible with variable range hopping~\cite{Han2010,Droescher2011,Bischoff2012} were observed. It is generally believed that the disorder responsible for the localization of charge carriers in graphene nanodevices  originates from rough edges, fabrication residues, substrate or the random orientation of the lattice. While there is some evidence that disorder originating from the edges plays an important role for transport~\cite{Bischoff2012,Engels2013}, the microscopic details and mechanisms are so far not well understood. To complement the experimental findings, there is a variety of theoretical work on graphene nanoribbons taking into account various defect and disorder scenarios~\cite{Supplementary}. As it is generally difficult to obtain information about graphene nanodevices on an atomic scale and to perform transport experiments for the same device, it was so far generally not possible to directly compare experiments and theory on a microscopic level. 

The electrical transport measurements of short and narrow single layer graphene constrictions on a hexagonal boron nitride substrate presented in this paper allow for a comparison of theory and experiment. From our measured data we develop a model for the shape of the envelopes of wave functions describing localized charge carriers inside the constrictions. We compare this model to tight-binding simulations of constrictions with similar geometry and non-perfect edges: our combined experimental and theoretical analysis suggests that the charge carriers are most likely localized along the rough edge of the constriction and, surprisingly, extend along the edge quite far out into the leads. We conclude that irrespective of the microscopic details of the edge disorder, its presence leads to localization of the wave functions along the edge on a length scale much shorter than the bulk localization length, but larger than the typical length scale of the physical edge disorder. We finally use these insigths to discuss previous experiments.

\section*{Experimental methods}
Our graphene constrictions have a length and width of about 30$\,\mathrm{nm}$ and are connected to wide graphene leads. The devices are shown in Figs.~1a,b and consist of a single-layer graphene flake on top of a hexagonal boron nitride flake (thickness 16$\,\mathrm{nm}$). The fabrication is similar to Ref.~\cite{Bischoff2012}. Device B is the same as device A with an additional lithography, evaporation and annealing step to increase the size of the metal contacts and therefore decrease the length $L$ of the wide graphene leads of the devices (the same is true for devices D and C). In addition to the constrictions, two micron-sized graphene stripes originating from the same graphene flake are located on the chip. They exhibit broken symmetry states in the quantum Hall regime confirming single-layer behavior as well as high material quality. All measurements shown in this paper were recorded at a temperature of 1.3$\,\mathrm{K}$ and with $+V_{SD}/2$ applied to the source and $-V_{SD}/2$ applied to the drain contact. In the whole investigated parameter range no leakage current from the back-gate to the devices could be observed. The resistances of the involved contacts were determined prior to etching the graphene and found to be well below 1$\,\mathrm{k\Omega}$.

\section*{Results}

\begin{figure*}[tbp]
	\centering\includegraphics[width=0.9\textwidth]{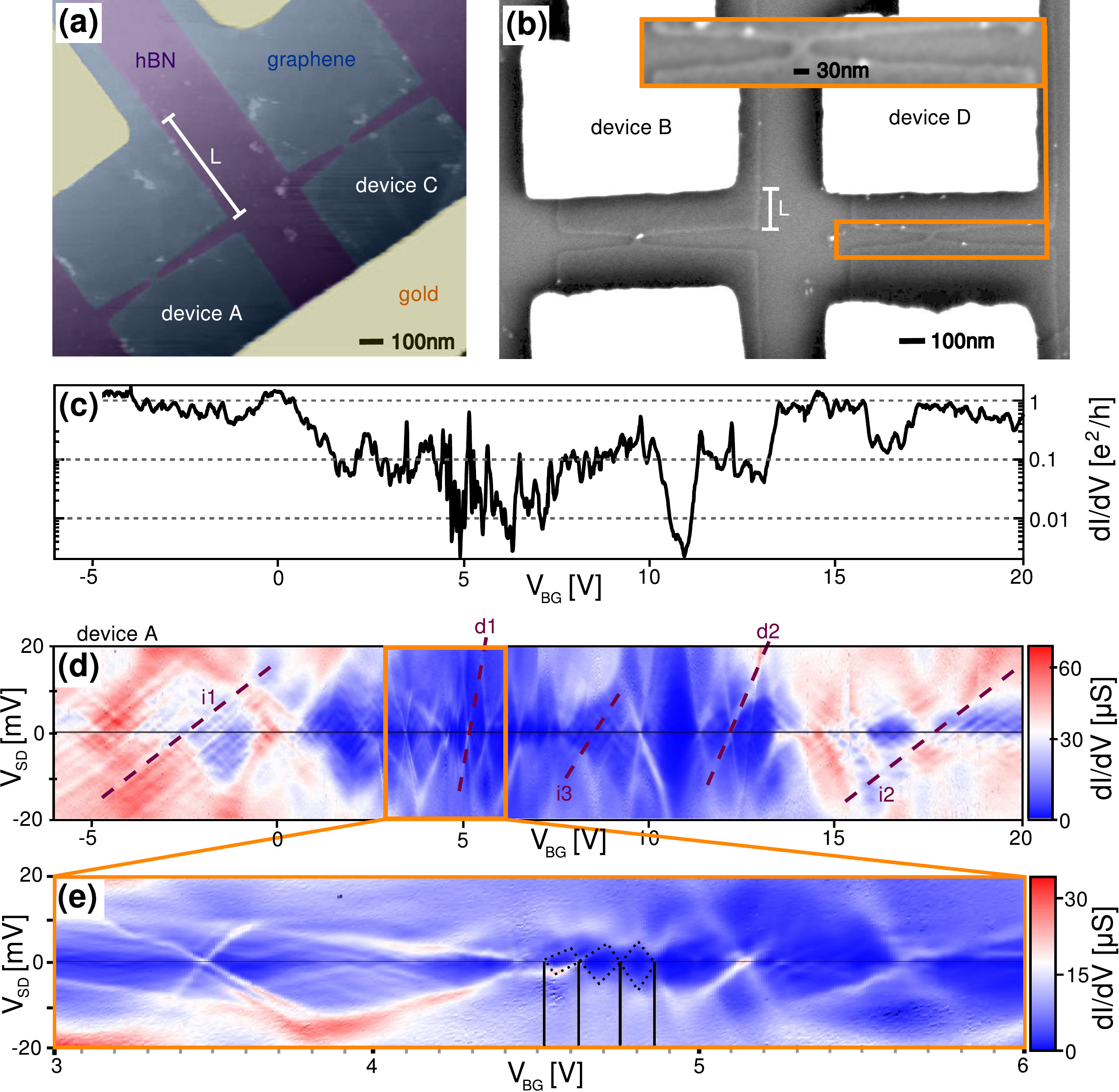}
	\caption{(color online) (a) Scanning force microscopy image (false color) of the two devices A and C. (b) Scanning electron microscopy image of the two devices B and D. (c) Differential conductance in logarithmic scale as a function of back-gate voltage for device A at zero applied bias. (d) Differential conductance of device A as a function of applied back-gate and bias voltage. (e) Close-up of a region of Fig.1d. }
\end{figure*}

\begin{figure*}[tbp]
	\centering\includegraphics[width=0.9\textwidth]{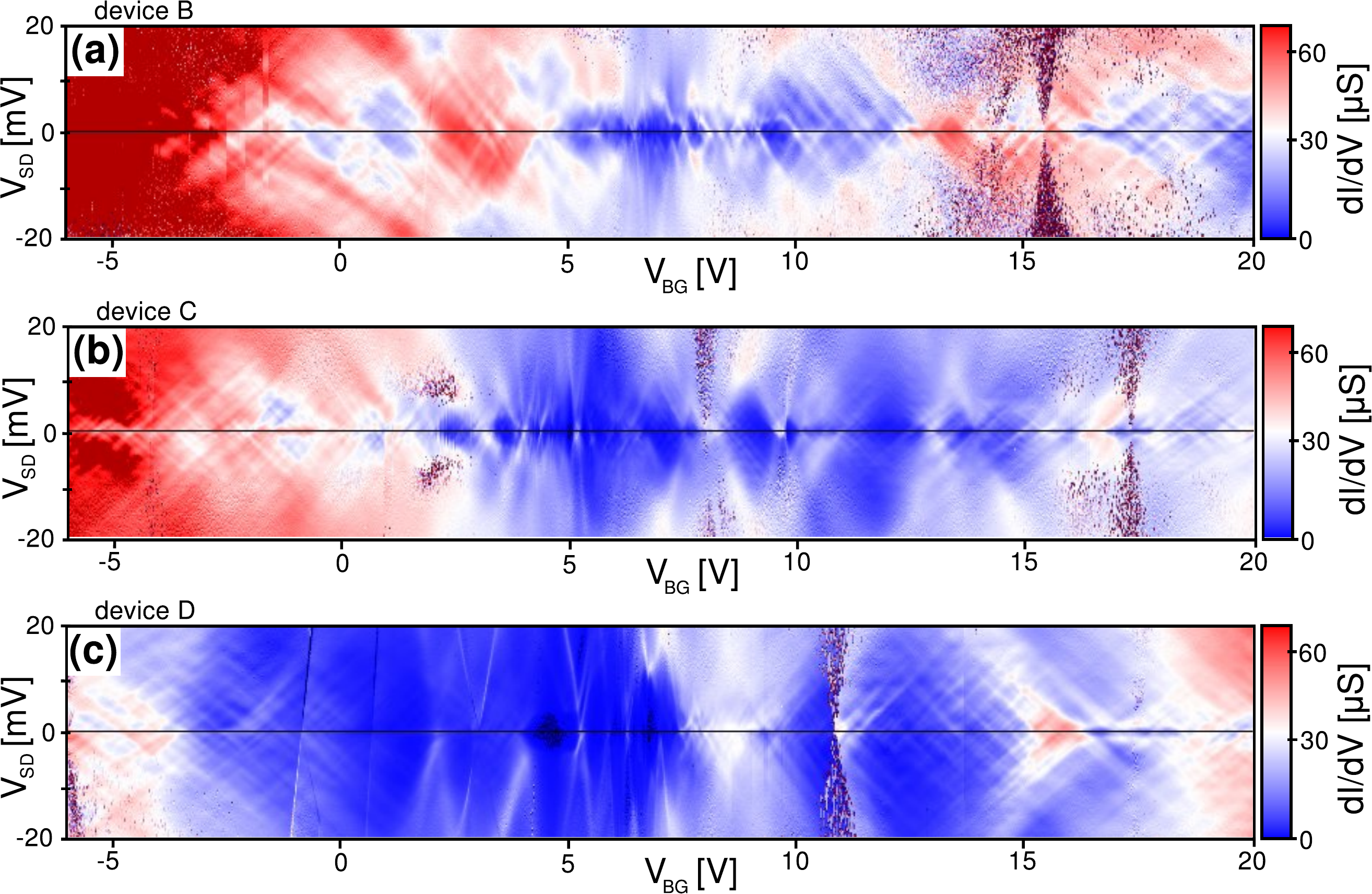}
	\caption{(color online) (a) Differential conductance as a function of applied back-gate and bias voltage for device B, (b) for device C, (c) for device D.}
\end{figure*}

\subsection*{Transport regime of the leads / bulk}
In order to separate influences in the conductance from the constriction and the wide graphene leads adjacent to it, we first estimate important transport length scales of the leads. As they cannot easily be measured on the device itself, we extract the values from one of the micron-sized graphene stripes. The disorder density extracted following Ref.~\cite{Du2008} is found to be lower than $10^{10}\,\mathrm{cm^{-2}}$. A lower bound of the hole mobility of 50'000$\,\mathrm{cm^{2}V^{-1}s^{-1}}$ and a lower bound for the electron mobility of 25'000$\,\mathrm{cm^{2}V^{-1}s^{-1}}$ at high carrier densities are determined, both approaching 100'000$\,\mathrm{cm^{2}V^{-1}s^{-1}}$ close to the Dirac point. The mean free path $l_e$ is of the order of 100$\,\mathrm{nm}$ close to the Dirac point and increases for higher carrier densities. The graphene leads will therefore likely be in a quasi-ballistic transport regime. The product $k_F l_e$ close to the Dirac point ($k_F$ is the Fermi wave-vector) is found to be slightly larger than one and to increase by about one order of magnitude for a 5$\,\mathrm{V}$ change in back-gate voltage~\cite{Supplementary}. This indicates that the charge carriers in the leads are not expected to be strongly localized (Ioffe-Regel criterion~\cite{Ioffe1960}). This is also visible in the estimated localization length $\xi\approx l_e \times \exp(k_F l_e / 2)$ which rapidly increases away from the Dirac point reaching microns at a density of $0.7\times 10^{11}\,\mathrm{cm^{-2}}$ (about 1$\,\mathrm{V}$ in back-gate away from the Dirac point).

\subsection*{Blockade regime, zero magnetic field}
In the following, the graphene constriction devices \mbox{A-D} are discussed. Fig.~1c shows the differential conductance at zero applied DC-bias as a function of back-gate voltage for device A. In the center region (i.e. around $V_{BG}=2\ldots 12\,\mathrm{V}$), the conductance is strongly suppressed ($\ll e^2/h$) whereas in the outer regions the conductance is about $e^2/h$. As discussed before, the strong suppression was not observed for the conductance of the leads alone and must therefore be attributed to the constriction.

We first discuss the region of suppressed conductance: when a finite DC-bias is applied, Coulomb-blockade diamonds are observed as shown in Fig.~1d (see Fig.~2 for the other devices). Coulomb-blockade diamonds in graphene nanoribbons were often studied in the  literature~\cite{Liu2009,Molitor2009,Stampfer2009b,Todd2009,Bai2010,Gallagher2010,Han2010,Molitor2010,Oostinga2010,Candini2011,Droescher2011,Terres2011,Wang2011,Bischoff2012,Ki2012,Pascher2012}. Before discussing details, we compare the values for the so-called ''source-drain-gap'' (region of suppressed conductance in source-drain voltage) and the so-called ''back-gate-gap'' (region of suppressed conductance in back-gate voltage) with values from the literature. 

We observe a large range of source-drain-gap values for our four different but nominally similarly sized devices A-D. We find a souce-drain-gap larger than 20$\,\mathrm{meV}$ for device D whereas the source-drain-gap of device B is only a few $\mathrm{meV}$. The back-gate-gaps are of the order of $5\ldots 10\,\mathrm{V}$ for all four devices. 

We limit our comparison to ribbons of about 30$\,\mathrm{nm}$ width that are relatively short (for long ribbons of 30$\,\mathrm{nm}$ width see Refs.~\cite{Han2007,Gallagher2010,Terres2011}). Molitor et al.~\cite{Molitor2009,Molitor2010} report measurements of 30$\,\mathrm{nm}$ wide and 100$\,\mathrm{nm}$ long ribbons and extract a back-gate-gap of the order of 25$\,\mathrm{V}$ and a source-drain-gap of the order of 35$\,\mathrm{meV}$. Todd et al.~\cite{Todd2009} show data of a 35$\,\mathrm{nm}$ wide ribbon with a length of 60$\,\mathrm{nm}$ where conductance never drops below about $e^2/h$ and therefore the back-gate-gap and the source-drain-gap are both zero. Gallagher et al.~\cite{Gallagher2010} show data for a 30$\,\mathrm{nm}$ long and 40$\,\mathrm{nm}$ long ribbon, where the source-drain-gap is well below 10$\,\mathrm{meV}$ and the back-gate-gap is a few $\mathrm{V}$. 

When comparing these values for different devices, special care needs to be taken as the capacitance per area to the back-gate will depend on the device geometry and therefore influence the width of the back-gate-gap. Also the source-drain-gap is not straight forward to compare as it is dominated by the capacitive coupling to neighboring leads and sites of localized charge (compare Refs.~\cite{Bischoff2012,Bischoff2013}). Therefore the source-drain-gap depends strongly on the exact arrangement of localized charge sites and on their number (compare e.g. Ref.~\cite{Wiel2003}). 

We conclude that the large variation of values found in the literature and in our measurements indicates that microscopic details play an important role for transport in such devices. We further conclude that both the source-drain-gap and the back-gate-gap should not be direcly compared for such short ribbons and that the values we observe are compatible with some of the values found in the literature. In order to add new insights to the already existing literature, we in the following carefully extract and discuss various details from our transport measurements.

First we focus on the region of suppressed conductance: the size of the diamonds in bias as well as the slopes (compare e.g. lines d1 and d2 in Fig.~1d) vary strongly even for neighboring diamonds. The change in slope of the diamonds indicates that the capacitive coupling between leads and localized charge varies strongly between different (neighboring) diamonds. The changing size in back-gate indicates a change in capacitance to the back-gate and therefore either a substantial change in size or position of the site on which charge is localized. Further, one can observe that inside many of the diamonds, current is not fully suppressed and that some fine structure is visible. This indicates that either several spatially parallel channels are available for electrons to flow or that co-tunneling processes are important (compare also Ref.~\cite{Bischoff2013}). In order to estimate the strength of the tunneling coupling, we identify one of the narrowest Coulomb-blockade peaks and find that it is broadened by tunneling rather than temperature~\cite{Supplementary}. We therefore argue that due to the high tunneling coupling, co-tunneling processes will definitely be important.

We estimate the area on which charge is localized: the smallest observed diamonds (see Fig.~1e) span about $0.1\ldots 0.2\,\mathrm{V}$ in back-gate voltage which corresponds to a capacitance of $1.6 \ldots 0.8\,\mathrm{aF}$ between the site of localized charge and the back-gate ($C_{BG,loc} = e/\Delta V_{BG}$). Based on the geometry of our constriction we expect an enhancement of capacitance between the constriction and the back-gate compared to a plate capacitor model due to stray fields at the edge of the device. This enhancement is expected to be rather small as most stray field lines are  screened by the graphene leads. Electrostatic simulations of our device geometry show an enhancement of less than a factor of 1.5 of the capacitance per area relative to a plate capacitor model for the constriction region of the device. We will later also justify this value based on data recorded in magnetic field. Employing the plate capacitor model corrected with a factor of 1.5 for stray fields~\cite{Supplementary}, the area of a site of localized charge can be calculated based on the relation $A \approx \frac{ed}{\epsilon\epsilon_0\Delta V_{BG}}\times\frac{1}{1.5} \approx 1 \mathrm{V}/\Delta V_{BG}\times (30\,\mathrm{nm})^2$. Consequently, the area associated with every Coulomb-blockade diamond that spans less than 1 V in back-gate is larger than the area of the constriction which is about $(30\,\mathrm{nm})^2$. The area associated with a diamond spanning 0.1$\,\mathrm{V}$ in back-gate for example is estimated to be 10 times larger than the geometrical constriction size. In this discussion we neglect the quantum capacitance of the graphene constriction as we expect its impact to be small~\cite{Reiter2014}. We also neglect a possibly existing confinement energy term as this would lead to a further increase the estimated areas.

Another value of interest is the coupling capacitance of a site of localized charge to the leads: assuming for the moment that only one site of localized charge is present, we can estimate from the charging energy the self-capacitance of this site of localized charge: $C^\Sigma = e/\Delta V_{sd}$, where $\Delta V_{sd}$ is half the size of the diamond in source-drain direction. In case of a single site of localized charge, the largest contributions to $C^\Sigma$ will be the coupling to source, drain and back-gate. For $\Delta V_{sd}=2\,\mathrm{mV}$ we get: $C^\Sigma \approx C_{s,loc} + C_{d,loc} + C_{BG,loc} \approx 80\,\mathrm{aF}$. The coupling capacitances to source and drain are therefore about a factor of 100 larger than the capacitive coupling to the back-gate and therefore determine the charging energy. In case of multiple sites of localized charge, the self-capacitance will be distributed between the leads and the other sites of localized charge and enhanced as discussed e.g. in Ref.~\cite{Wiel2003}. In the above estimate we neglected the influence of a possibly existing quantum confinement term that might contribute to the height of the Coulomb-blockade diamond and would therefore increase the value for the estimated self-capacitance even further.

\subsection*{Interference regime, zero magnetic field}

In the regime in Fig.~1d where the resistance at zero DC source-drain voltage is in the range of about $10\ldots20\,\mathrm{k}\Omega$, we observe many parallel lines (see lines i1, i2) with a spacing of about $0.3\ldots0.6\,\mathrm{V}$ in back-gate voltage. Since this checkerboard-pattern is also visibly superimposed on top of some of the diamonds (see line i3 in Fig.~1d) and disappears at a temperature of about 12$\,\mathrm{K}$ (whereas the Coulomb-blockade diamonds survive up to about 50$\,\mathrm{K}$), we conclude that it is due to the conductance of the graphene leads rather than the constriction. In the next section we will further justify this conclusion by analyzing the B-field dependence of this pattern.

Such a checkerboard pattern is generally attributed to phase-coherent interferences and was previously observed in graphene nanoribbons~\cite{Gallagher2010,Todd2009}, in bulk graphene~\cite{Miao2007} and a variety of quasi-1D systems~\cite{Liang2001,GroveRasmussen2007,Kretinin}. In the case of quasi-1D systems, the checkerboard pattern can be interpreted as Fabry-Pérot interferences between two contacts. The tilt of the resonance lines is attributed to the back-gate changing the position of the Fermi energy and therefore changing the wave-vector of the different modes~\cite{Liang2001}. In the devices investigated in this paper, we do not expect a simple Fabry-Pérot pattern due to the geometry. For the observed oscillation amplitude of about $0.3\,e^2/h$, we estimate a phase coherence length of the order of a few hundred nm~\cite{Supplementary} which is compatible with previous experiments~\cite{Russo2008,Huefner2010}.

\begin{figure*}[htbp]
	\centering\includegraphics[width=0.9\textwidth]{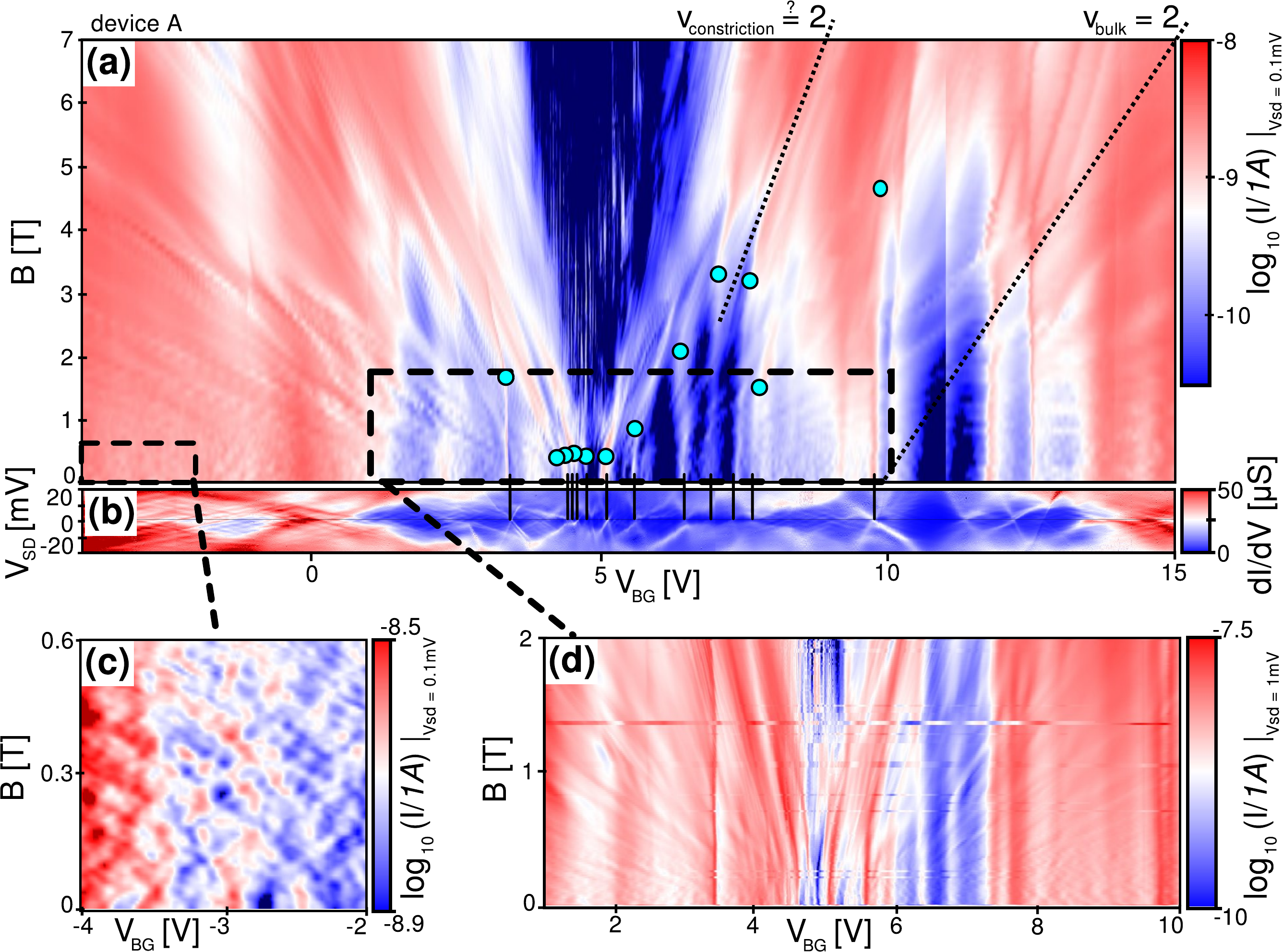}
	\caption{(color online) (a) Current flowing through device A as a function of back-gate voltage and perpendicular magnetic field at a fixed bias of 0.1mV. Cyan circles mark the approximate position at which the Coulomb-blockade peaks start to tilt. Black dashed lines mark the slope of the $\nu=2$ plateau for a micron-sized device and as a guide to the eye the predominant slope observed in this measurement. (b) Same plot as Fig.~1c added for direct comparison. (c,d) Zooms into different regions of Fig.~2a.}
\end{figure*}

\subsection*{Finite magnetic field}

Fig.~3a shows the current flowing through device A as a function of perpendicular magnetic field and back-gate voltage for a fixed source-drain bias of 0.1$\,\mathrm{mV}$. The general shape of the observed features does not change with moderately higher bias, but the visibility of small features decreases. The plot can be divided into four different regimes:

(i) \textit{Low magnetic field, blockade regime:} in this regime (see also zoom in Fig.~3d), the Coulomb-blockade diamonds are unaffected by the magnetic field (compare Fig.~3a with Fig.~3b -- black solid lines mark exemplarily some corresponding Coulomb-blockade resonances in both plots). Coulomb-blockade peaks stay unchanged in magnetic field until the lines start to bend at a certain field value. The approximate points where the lines bend are marked by light blue markers in Fig.~3a. From the value in B-field where this happens, the length scale important for transport can be estimated: Coulomb-blockade is expected to dominate as long as the magnetic length~\cite{Morimoto2008} $l_m = \sqrt{\hbar/eB}$ is larger than the length scale on which charge is localized. For B=0.5$\,\mathrm{T}$ the magnetic length is $l_m\approx 35~\mathrm{nm}$ comparable to the constriction dimensions.

(ii) \textit{Low magnetic field, outside of blockade:} this is the regime where an interference pattern is observed (see also zoom in Fig.~3c). Transport changes on the order of a few ten mT which corresponds to an area ($A\approx \frac{h}{e}\frac{1}{\Delta B}$) of a few hundred nanometers square. Also the magnetic field scale on which transport changes is increased for devices B and D indicating a smaller relevant area. This confirms that the interference pattern indeed arises from the leads and not from the constriction. These measurements can again be used to estimate the phase coherence length and we find similar values as before~\cite{Supplementary}.

(iii) \textit{High magnetic field, away from the Dirac point:} in this regime, different sets of approximately parallel lines are visible. Most of these lines anti-cross with each other. In previous experiments such lines have been observed and were attributed to localized states in the quantum Hall regime~\cite{Martin2009,Lee2012,Velasco2010}. Many of these lines show a slope which is steeper than that of the dominating $\nu=\pm 2$ filling factor in the leads. As those lines start to appear at quite low magnetic field it is unlikely that they belong to the broken symmetry state $\nu=\pm 1$ of the leads. The most probable explanation is that these lines originate from the $\nu=\pm 2$ filling factor in the constriction. This allows us to determine the average charge density in the constriction and therefore extract an enhancement factor of roughly 1.8 of the plate capacitor model. This value is slightly larger than the value estimated at zero magnetic field with electrostatic simulations.

(iv) \textit{High magnetic field, around Dirac point:} in this regime, current is heavily suppressed. For micron-sized graphene stripes fabricated from the same graphene flake, the $\nu=0$ broken symmetry quantum Hall state starts to emerge at high magnetic fields. The strong suppression of conductance in the graphene leads together with the constriction is likely responsible for the observed triangular shape of strongly suppressed conductance. Additional sharp lines are visible that do not show any B-field dependence. This indicates that in this regime charge is localized on length scales of a few tens of nanometers.

This behavior described in (i)-(iv) is also observed for the other three devices~\cite{Supplementary}.

\section*{Interpretation of the Results}

\subsection*{Wave function localized along the edge}
The observation of Coulomb-blockade requires that charge carriers are localized on a finite area of the device and that the wave function of these localized charges exhibits a small coupling to the wave functions of both leads. Similarly one can interpret a Coulomb-blockade peak as a scattering state that can be decomposed into incoming / outgoing (delocalized) scattering channels and a localized eigenstate of the constriction (similar to a Feshbach decomposition from scattering theory~\cite{Feschbach}). While the exact spatial envelope of these localized states depends on the microscopic details of the constriction, we identify their common properties based on our experimental results from various devices: 
\begin{enumerate}
	\item their effective area varies strongly in size. Remarkably, several states are substantially larger than the constriction,
	\item they need to couple weakly to both leads to result in Coulomb-blockade, and
	\item consecutive states in energy strongly vary in effective areas and coupling strength to the leads.
\end{enumerate}

Most calculations so far focus on ribbon geometries only and do not consider the adjacent graphene leads. Sols et al.~\cite{Sols2007} suggested that the roughness introduced by the etching process separates the nanoribbon into islands where Coulomb-blockade appears. Along similar lines, Evaldsson et al.~\cite{Evaldsson2008} show theoretically that edge defects locally deplete the density of states in the ribbon, forming an effective barrier trapping charge carriers. Several other researchers have also discussed edge-disorder induced Anderson localization~\cite{Gunlycke2007,Evaldsson2008,Querlioz2008,Mucciolo2009,Xu2009,Libisch2012}. It is not clear whether these results apply to our geometry which features extremely wide (bulk) leads. Other models predict localization in the bulk based on suppressed Klein tunneling between disorder-induced puddles~\cite{Stampfer2009b}. However, our experimental evidence suggests that the range in back-gate voltage where puddles are expected (disorder density) is significantly smaller than the range over which Coulomb-blockade is observed. Furthermore, previous investigations do not find a significant role of bulk disorder in etched graphene nanostructures on hexagonal boron nitride~\cite{Bischoff2012}. 

As a scattering state localized only in the constriction (see Fig.~4a) fails to fulfill our first criterion, the most obvious solution would be spatially extended states it into the bulk of the graphene (see Fig.~4b). Such a state is however not consistent with our measurements since it would either require strong bulk disorder for sufficient localization or lead to a loss of Coulomb-blockade due to strong wave function overlap. 

Instead of being localized in the bulk, the wave function could be primarily localized along (or close to) the edges of the constriction extending into the leads along the edges (see Figs.~4c,d). Such a shape limits the overlap between the quasi one-dimensional constriction state and the delocalized incoming and outgoing scattering states, resulting in the observed narrow Coulomb-blockade peaks. Also by being localized primarily at the edge, stray fields play an important role for the capacitance to the back-gate. We estimate that the length on which such a wave function is localized should be of the order of hundred nanometers as deduced from the narrowest diamonds~\cite{Supplementary}. Disorder at the edges of the constriction must thus lead to localization along the edge on a length scale significantly longer than the size of our constriction. We further note that for reactive-ion-etched devices, this localization length is much larger than the physical disorder length which is of the order of nanometers~\cite{Bischoff2011}. It is worth noting that these states of localized charge along the edge are different from edge states found in perfect zig-zag nanoribbons~\cite{Nakada1996}.

Infering such detailed properties of the Coulomb-blockade state in this experiment is only possible because the constriction is significantly smaller than the length on which the wave function is localized and because the geometry was carefully chosen to keep the capacitance between the constriction and the back-gate small.

\begin{figure}[tbp]
	\centering\includegraphics[width=\columnwidth]{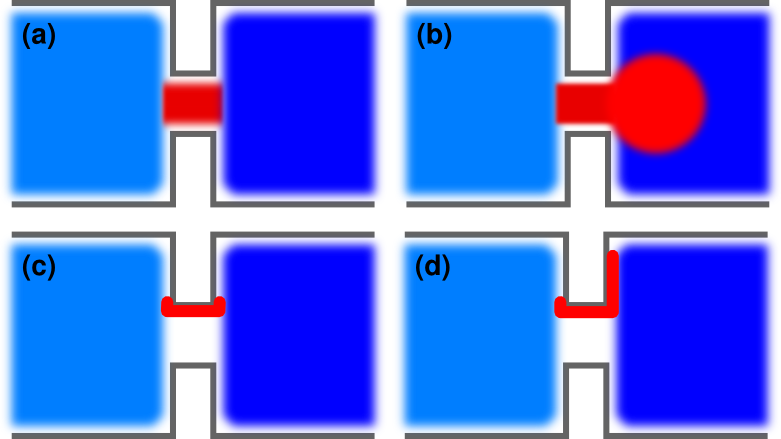}
	\caption{(color online) Schematic drawing of possible wave function envelopes (red) that are localized in the constriction and couple to the extended wave functions in the graphene leads (blue). Charge carriers are primarily localized (a) in the full area of the constriction, (b) in the constriction and the right lead, (c) along the edge of the constriction and (d) along the edge of the constriction and the right lead.}
\end{figure}

\subsection*{Simulation of the constrictions}
In order to better characterize the spatial extent of the localized part of the scattering state, we calculate the quasi-bound state by tight-binding simulations. We model a $30\times 30\,\mathrm{nm}$ constriction conntected to $140\,\mathrm{nm}$ wide leads. We use a third-nearest-neighbor tight-binding approach~\cite{Libisch09} as well as open boundary conditions to the left and right~\cite{Libisch2012} and rough edges forming the constriction at the top and bottom (see Fig.~5.). To simulate the experimental edge roughness we include random fluctuations of the boundary with an amplitude of $2\,\mathrm{nm}$~\cite{Supplementary}. 

Since we model resonant states of a scattering experimen as an open quantum system, we obtain complex eigenvalues whose imaginary parts describe the coupling strength to the leads. We find two distinct classes of eigenstates:%
\begin{itemize}
	\item delocalized states with strong coupling to the left and/or right leads (see Fig.~5a), featuring an imaginary part larger than the average level spacing
	\item states strongly localized at the device edges (see Fig.~5b-d) with imaginary parts much smaller than the average level spacing
\end{itemize}
As a suitable measure for their coupling to the leads of the structure (the imaginary part of the eigenenergies) and of the degree of localization we employ the inverse participation ratio (IPR)~\cite{Thouless}
\begin{equation}\label{eq:IPR}
\mathrm{IPR} = \frac{\langle \psi^4 \rangle}{\langle \psi^2 \rangle^2} 
\end{equation} 
where the brackets $\langle\ldots\rangle$ denote a spatial average over the whole device. $\mathrm{IPR} \gg 1$ for strongly localized states (the forth power dominates) and approaches one for a perfectly delocalized wave function. We find many localized states (large IPR) around the Dirac point (see Fig.~5e). We further distinguish localized states by their amplitude in the area $\mathcal{A}_C$ of the constriction,
\begin{equation}\label{eq:P}
P = \int_{\mathcal{A}_C}\left|\psi\right|^2 \mathrm{d}^3r.
\end{equation}
States localized mostly in the constriction ($P > 0.75$, see blue triangles in Fig.~5e) will contribute to Coulomb-blockade and are likely to result in wide Coulomb-blockade diamonds. We also find a number of strongly localized states that localize along the rough edge of the constriction and extend along the edge also into the lead parts of the device (see red squares in Fig.~5e). We conjecture that these states result in the smaller diamonds observed in experiment. While we cannot expect to quantitatively simulate the disorder observed in experiment, we can reproduce several features observed in experiment, including:
\begin{itemize}
	\item a strong variation in effective area, or IPR, for states localized in the constriction
	\item an energy window of about 80 meV around the Dirac point where localized states dominate and Coulomb-blockade should occur
	\item states with effective areas of the wave functions much larger than the constriction size that, nevertheless, show only weak coupling to the leads and relatively high amplitude at the constriction. They should give rise to Coulomb-blockade diamonds.
\end{itemize}
We note that the localization length along the rough edge strongly depends on the microscopic details of the edge roughness~\cite{Libisch2012}. Our simulations qualitatively reproduce all experimental observations. The very large parameter space for possible edge roughness configurations (including possible lattice defects, adsorbates, and geometrical edge variations) calls the merit of a quantitative comparison with experiment into question.

\begin{figure}
	\includegraphics[width=\columnwidth]{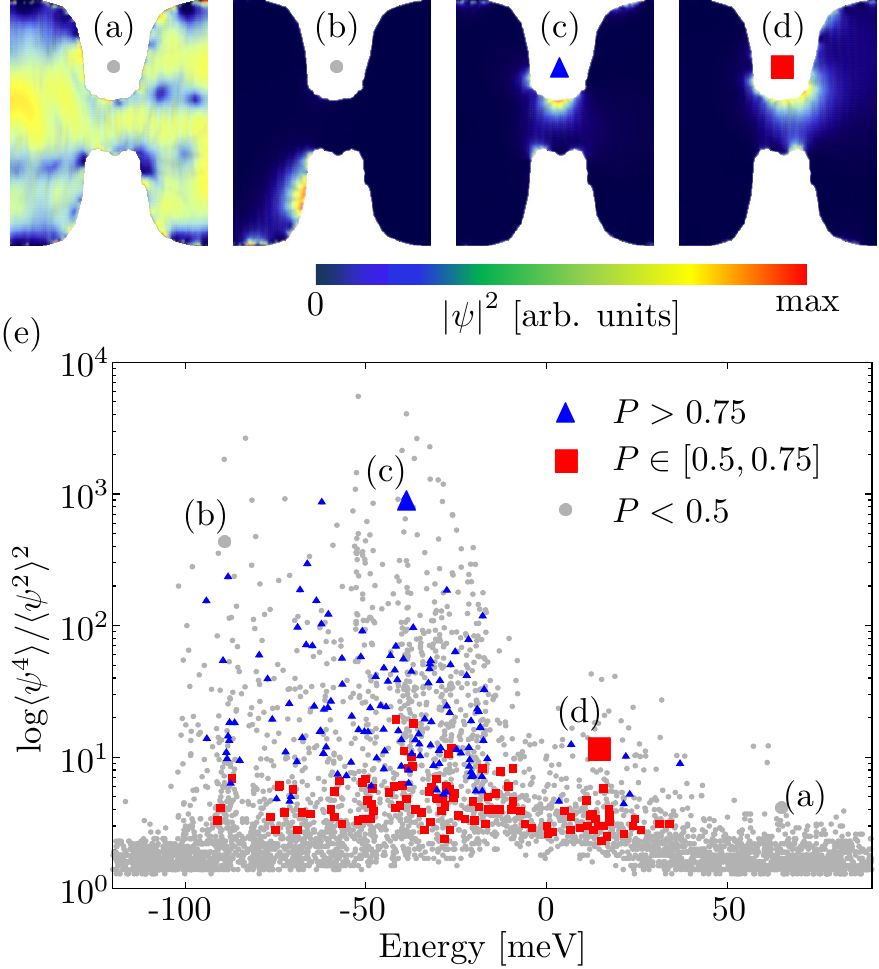}
	\caption{(color online) Results for the tight-binding simulation of a $30\times30\,\mathrm{nm}$ graphene nanoconstriction connected to open leads of width $140\,\mathrm{nm}$, for different microscopic realizations of edge disorder. (a)-(d) Four eigenstates of the nanoconstrictions [see symbols in (e)], (e) Statistics of the IPR [Eq.~(\ref{eq:IPR})] of 5000 eigenstates as a function of eigenenergy. States with large probability $P$ [Eq.~(\ref{eq:P})] to be inside the nanoconstriction are marked by blue triangles ($P > 0.75$) or red squares ($P \in [0.5,0.75]$, see inset). Larger symbols denote the four states depicted in (a)-(d).}
\end{figure}

\subsection*{Insights into previous experiments}
In this final section, we apply these new findings to previous experiments to check for discrepancies and to better understand transport in graphene nanodevices. While this section is rather speculative, we find it useful to formulate open questions that need to be addressed in future experiments. 

Quite surprisingly, graphene nanoribbons fabricated by different techniques in different laboratories most of the time exhibited quite similar transport properties (for comparable device sizes and within the large variations between different devices fabricated in the same run). Both qualitative behavior as well as quantitative values are reproduced in different experiments (compare e.g. Refs.~\cite{Han2007,Molitor2009,Stampfer2009b,Todd2009,Bai2010,Gallagher2010,Han2010,Molitor2010,Oostinga2010,Droescher2011,Terres2011,Hwang2012,Smith2013} for low-temperature  measurements). These devices will likely differ in crystallographic edge orientation, edge roughness, edge termination, fabrication residues, geometry and various other aspects. On the other hand, the details usually differ for different cooldowns of the same device (compare e.g. Ref.~\cite{Bischoff2012}). Together with the findings from this paper, this leads to the following hypothesis: the presence of a certain amount of disorder at the edges is sufficient to localize charge carriers close to the edges. The details of the edge (and likely also of the bulk) determine the envelope and energy of such localized wave functions. As long as disorder is weak enough that the wave function is not localized on the same length scale as the disorder, the details of the disorder seem to be unimportant for the qualitative picture. 

While edge disorder is so far hard to control technologically, area disorder is more easily accessible experimentally: in one experiment, bulk disorder was reduced and no significant change in transport was observed~\cite{Bischoff2012}. In another experiment, disorder was increased by deposition of single atoms onto the ribbon: the higher the amount of additional atoms on top, the more suppressed the transport~\cite{Smith2013}. This might indicate that in the first regime, area disorder was sufficiently low such that further reducing it did not change transport whereas in the other case the additional atoms on top of the graphene provided sufficiently strong disorder such that the wave function could now also be localized inside the ribbon. 

Since our findings suggest that edge disorder plays a crucial role in determining device properties, we can also qualitatively understand why wider ribbons ($w>100\,\mathrm{nm}$) do not usually feature Coulomb-blockade~\cite{Han2007,Molitor2009,Han2010,Molitor2010,Terres2011}: the small edge-to-bulk ratio for wider ribbons diminishes the influence of the rough edges on overall transport. Localization along the edge still happens but transport is dominated by bulk contributions. 

For graphene ring experiments, it was found that the area on which transport happens is generally smaller than the ring width~\cite{Russo2008,Huefner2010}. Disordered edges that localize electrons would therefore explain this spatially reduced width.

A further striking difference between devices etched in ribbon geometry and in island geometry (quantum dot) is that the former ones usually display chaotic Coulomb-blockade whereas the latter ones often display quite regular Coulomb-blockade diamonds. As discussed in detail in a previous experiment~\cite{Bischoff2013}, the observation of these regular and often non-overlapping Coulomb-blockade diamonds is due to an arrangement of three sites of localization in series together with higher order co-tunneling. The present findings suggest that the regular diamonds occur due to the formation of a quantum dot in each constriction. Consequently, the island is decoupled from the leads and therefore will form an additional quantum dot. The wave function will likely be distributed over the two edges of the island which have to be coupled in order to be compatible with the experiments. A more detailed discussion of possible wave function envelopes for island geometries together with tight-binding calculations is provided in the supplementary materials~\cite{Supplementary}.

\section*{Summary and Conclusion}
We have shown that short and narrow graphene constrictions on a hexagonal boron nitride substrate still show Coulomb-blockade at low temperature in agreement with previous work. By carefully analyzing the capacitances corresponding to the observed Coulomb-blockade diamonds, we found that our results are incompatible with any model resulting primarily in charge localization  in the bulk of the ribbon. Our experiments can be explained by states that are mostly localized along the highly disordered edge of the graphene. This localization can extend along the device edge into the leads. While these experimental findings improve our understanding of transport in graphene nanostructures significantly, the microscopic details of the edges are still not well understood. We therefore suggest further experiments along two different routes: first, similar measurements should be performed with different edge morphology obtained either by chemical passivation or different fabrication techniques. A second and much harder route is performing scanning tunneling experiments with atomic resolution combined with transport experiments: this would allow to get a microscopic understanding of typical edge configurations and might potentially also allow to locally alter the edge and probe the local density of states.

\section*{Acknowledgements}
Financial support by the National Center of Competence in Research on “Quantum Science and Technology“ (NCCR QSIT) funded by the Swiss National Science Foundation is gratefully acknowledged.  FL and JB acknowledge support by the SFB-041 ViCoM. We thank Pascal Butti for performing the electrostatic simulations of the capacitance. We further thank Anastasia Varlet, Pauline Simonet, Fabrizio Nichele, Nikola Pascher, Sarah Hellmüller, Clemens Rössler and Annina Moser for helpful discussions.



\end{document}